\documentclass[twocolumn,aps,prb,amssymb,raggedbottom,nobalancelastpage,superscriptaddress]{revtex4-1}

\usepackage{amsmath}
\usepackage{amssymb}
\usepackage{amsfonts}
\usepackage{dsfont}
\usepackage{graphicx}
\usepackage{bm}
\usepackage{color}
\usepackage{appendix}
\usepackage{epsfig}

\def \etal{{\it et al.}~}

\begin{document}
\title{Comment on ``Topological equivalence of crystal and quasicrystal band structures''}

\author{Yaacov E.~Kraus}
\affiliation{Department of Condensed Matter Physics, Weizmann Institute of Science, Rehovot, Israel.} %
\author{Zohar Ringel}
\affiliation{Department of Condensed Matter Physics, Weizmann Institute of Science, Rehovot, Israel.} %
\author{Oded Zilberberg}
\affiliation{Theoretische Physik, Wolfgang-Pauli-Strasse 27, ETH Zurich, CH-8093 Zurich, Switzerland.} %

\begin{abstract}
Madsen \etal [arXiv:1307.2577] claim that one-dimensional insulating crystals and one-dimensional insulating quasicrystals are topologically equivalent and, thus, trivial. In this comment, we clarify that in topological classification of one-dimensional systems, quasiperiodicity plays a role similar to that of a symmetry. Hence, periodic systems are out of the quasiperiodic topological space, and the equivalence between quasicrystals and crystals becomes immaterial.
\end{abstract}


\maketitle

In a recent paper~\cite{Madsen}, Madsen \etal argue that ``there is no topological distinction between one-dimensional (1D) insulating crystals and 1D insulating quasicrystals''. This statement seemingly contradicts an earlier work by the authors of this comment and collaborators~\cite{us1}, which showed that 1D quasiperiodic (QP) Hamiltonians may be topologically non-trivial, even in the absence of symmetry; whereas similar periodic Hamiltonians are known to be topologically trivial~\cite{LudwigDim}. We detail below that this alleged contradiction is due to the facts that in Ref.~[\onlinecite{Madsen}] (i) there is an attempt to apply topological considerations beyond the range of their validity, and (ii) the analysis relies on boundary states where they are not supposed to exist. In fact, the numerical results of Ref.~[\onlinecite{Madsen}] are in no contradiction with the claims of Ref.~[\onlinecite{us1}].

At first, let us repeat in detail our statement, appearing in Refs.~[\onlinecite{us1,us2,us3}]. We consider 1D noninteracting tight-binding models with QP terms, i.e.~terms of the form $f(n) c^\dagger_n c_{n'}$, where $c^\dagger_n$ creates a particle at site $n$, $n'$ is a site at the vicinity of site $n$, and $f$ is a QP function. A function $f(x)$ is QP if $f(x) = F(2\pi b_0 x,2\pi b_1 x +\phi, ..., 2\pi b_N x+\phi)$, where $F(x_0,x_1,\ldots,x_N)$ is a function of $N+1$ variables that is $2\pi$-periodic in all of them, and the $b_i$'s are mutually irrational~\cite{Moser1967,Hale1969}. The modulation of the underlying lattice is expressed by $b_0$, which we set to be $b_0 = 1$. The parameter $\phi$ shifts the QP modulations with respect to the underlying lattice. A simpler example is the Harper model, where the QP term is $\cos( 2\pi b_1 n + \phi ) c^\dagger_n c_n$, namely, $n' = n$, $N = 1$ and $F(x_0,x_1) = \cos(x_0)\cos(x_1)$. For simplicity, in the following we assume $N = 1$, and denote $b_1$ by $b$.

For any such QP Hamiltonian, each bulk gap can be associated with an integer Chern number~\cite{us1}. A Chern number associated with a gap can change its value only by closing the bulk gap or by exiting the above space of QP Hamiltonians. This association relies on the fact that in the QP terms, the effect of shifts in the parameter $\phi$ is physically indistinguishable from that of lattice translations. Therefore, bulk properties of QP Hamiltonians, such as the energy spectrum, do not depend on $\phi$. In order to associate an integer Chern number with a gap, one usually~\cite{Thouless_pump} has to integrate the Chern density over a period of $\phi$. However, since the Chern density is a bulk property, it does not depend on translations, and thus neither on $\phi$. Therefore, the integration over $\phi$ can be replaced by a constant multiplication by $2\pi$, and still results in an integer number. This way, each gap is associated with a Chern number for any given $\phi$-value. Notably, boundary phenomena, such as subgap boundary states, do depend on $\phi$ in QP systems.

Contrastively, in periodic Hamiltonians, where $b$ is rational, $\phi$ affects both bulk and boundary properties. Specifically, the Chern density is no longer independent on $\phi$. Hence, in order to associate the energy gaps with Chern numbers, the integration over $\phi$ is imperative, i.e.~only the topological pump may have a Chern number~\cite{Thouless_pump}. Hence, associating a Chern number to any single system with a given $\phi$ is impossible. Therefore, one can think of quasiperiodicity as a restriction which yields new topological phases, similar to the way time-reversal symmetry admits topological insulators.

The association of QP-models with Chern numbers has physical implications, such as: (i) Consider a Hamiltonian which embodies a parameter that continuously deforms between two 1D QP systems while maintaining the quasiperiodicity. For example, it contains two QP terms and a knob that continuously turns off one term and simultaneously turns on the other term. If the chemical potential is set to a gap whose Chern number changes its value between the two systems, then upon an adiabatic turn of the knob, there will be a point where the bulk gap closes. (ii) Place two semi-infinite QP systems with a different Chern number on a line adjacent to each other, and form a smooth interface region that interpolates between them. In such a case, subgap excitations will appear along the interface with an energy spacing that decays with the length of the interface. A similar phenomena has been experimentally observed in Ref.~[\onlinecite{us3}]. Note that, if the interface is sharp, the existence of subgap states is no longer guaranteed, since the sharp boundary locally breaks the quasiperiodicity. This is somewhat analogous to the absence of surface states in weak~\cite{us_WTI} and crystalline~\cite{Fu_CTI} topological insulators. (iii) Consider a semi-infinite QP system, where the chemical potential is set to a gap with a non-trivial Chern number. Now, measuring the local density of states (LDOS) at the boundary during a repeated etching of the last atomic site, a boundary state disappears and reappears. Notably, the measured energy of the boundary state gradually fills the bulk gap. In contrast, in periodic systems (i) topology does not prevent deformation between systems while keeping their bulk gap open, (ii) interface states would not fill the gap even for infinitely long interfaces, and (iii) the subgap LDOS at the etched boundary would be composed of only discrete energies and would not fill the bulk gap.

The above results are exact for infinite QP systems. Given a finite system, some discrepancies may appear, which vanish as the number of sites, $L$, goes to infinity. This is similar to all types of phase transitions. Furthermore, perturbation theory implies that a small perturbation to the Hamiltonian, which changes the matrix elements by less than $\epsilon \ll 1$, may open the gap during the phase transition to the order of $\epsilon$. Consequently, a system with $L$ sites and modulation frequency $b$ is similar to a system with modulation frequency $b' \in (b-\epsilon /L, b+\epsilon /L)$, up to corrections of order $\epsilon$. Therefore, as the system becomes shorter, the topological classification becomes blurred, as well as the distinction between rational and irrational modulations.

Turning back to infinite systems, a change of $b$ is no longer a small perturbation. It is always of the order of the unperturbed terms, since a change of $b$ to $b' = b + \epsilon$, causes intervals of the QP terms to flip their signs. Therefore, if a system with a \emph{rational} $b'$ is deformed into a system with an irrational $b$, all the aforementioned phenomena will appear, even for an extremely small $\epsilon$ (as long as $L$ is infinite). Complementary, if a system with an \emph{irrational} $b$ is deformed into a rational $b'=p/q$ system, discrepancies of the order of $1/q$ may appear~\cite{us1}. It then follows that given some uncertainty in $b$, one might have $1/q_\text{min}$-discrepancies, where $p_\text{min}/q_\text{min}$ is the rational number in the uncertainty interval with the smallest denominator. Such discrepancies accompanies all QP phenomena~\cite{QC_Janot}, e.g.~localization-delocalization transitions~\cite{AA}. Nevertheless, by choosing appropriate irrational $b$, such as the Golden ratio, imperfect experimental systems show signatures of both the topological~\cite{us3} and localization~\cite{Roati08,Lahini} behavior.

\vspace{5mm}

Madsen \etal~\cite{Madsen} consider the Aubry-Andr\'{e} model~\cite{AA}, following Ref.~[\onlinecite{us1}]. The spectrum of this model is gapped, and boundary states traverse its gaps as a function of $\phi$. The authors show numerically that the intervals of $\phi$, in which these states appear, strongly depend on the details of the boundary, and hence the model is seemingly topologically trivial. This dependence has been pointed out already in Ref.~[\onlinecite{us1}]. As explained above, the nontriviality of a 1D QP model does not imply the existence of subgap states at sharp boundaries for any given $\phi$, since the boundary breaks locally the quasiperiodicity. Therefore their absence does not prove their triviality.

Furthermore, the model was numerically diagonalized as a function of $b$, which was continuously varied from $1.60$ to $1.63$. Along this deformation the two largest spectral gaps remained open. One may think that it implies that all systems in this interval are adiabatically connected to each other. Thus, since the standard classification of band insulators~\cite{LudwigDim} classifies all gapped periodic 1D systems (in the absence of symmetry) as topological trivial, this should also be true for the QP systems. Seemingly, this flow of logic has led the authors of Ref.~[\onlinecite{Madsen}] to conclude that periodic and QP systems are topologically equivalent, and both are trivial band insulators.

We find this equivalence invalid, since it ignores the essential requirement for quasiperiodicity. As emphasized above, only the space of QP Hamiltonians is topologically classified by Chern numbers, whereas periodic Hamiltonians are not classified at all. Therefore, an open gap during a deformation from a QP into periodic system does not imply triviality in any sense. Intuitively, it can be thought of as arguing for the triviality of a topological insulator by adiabatically connecting it to an insulator with broken time-reversal symmetry~\cite{Nagaosa:2013}.

The fact that the two largest gaps remain open in the interval, $[1.6,1.63]$, nevertheless, implies that all the QP Hamiltonians in this interval indeed share the same Chern numbers in their largest gaps. Therefore, if the chemical potential of two such QP systems is set to one of these gaps, they are indeed topologically equivalent. It should be noted, however, that the Chern numbers of the small gaps change considerably when $b$ is changed.

Moreover, in the numerical data presented in Ref.~[\onlinecite{Madsen}], as $q_\text{min}=5$ in the interval $b \in [1.6,1.63]$, it is reasonable that the largest gaps persist in the whole interval. However, unlike irrational cases, for rational modulations in this interval we expect that these gaps and their corresponding Chern densities would fluctuate with $\phi$. This fluctuations becomes stronger as one approaches a phase transition, where the gaps diminish, and there is no longer a clearly defined Chern number for a given $\phi$. One might, however, suggest to define the Chern numbers of the aforementioned large gaps in the rational cases, simply by continuity. The drawback of such a definition is that it has no physical implications, i.e.~the aforementioned consequences (i)-(iii) would correspond to the periodic case, and thus the ``continued Chern number'' does not define a distinct state of matter.

To conclude, we see no contradiction between our theoretical claims~\cite{us1,us2,us3,us4} and the numerical results reported by Madsen \etal~\cite{Madsen}.

\vspace{5mm}

We thank A.~Stern, D.~Ivanov, and S.~Huber for fruitful discussions. %
We acknowledge the Minerva Foundation of the DFG, US-Israel BSF, and SNF for financial support.


\begin{thebibliography}{16}%
\makeatletter
\providecommand \@ifxundefined [1]{%
 \@ifx{#1\undefined}
}%
\providecommand \@ifnum [1]{%
 \ifnum #1\expandafter \@firstoftwo
 \else \expandafter \@secondoftwo
 \fi
}%
\providecommand \@ifx [1]{%
 \ifx #1\expandafter \@firstoftwo
 \else \expandafter \@secondoftwo
 \fi
}%
\providecommand \natexlab [1]{#1}%
\providecommand \enquote  [1]{``#1''}%
\providecommand \bibnamefont  [1]{#1}%
\providecommand \bibfnamefont [1]{#1}%
\providecommand \citenamefont [1]{#1}%
\providecommand \href@noop [0]{\@secondoftwo}%
\providecommand \href [0]{\begingroup \@sanitize@url \@href}%
\providecommand \@href[1]{\@@startlink{#1}\@@href}%
\providecommand \@@href[1]{\endgroup#1\@@endlink}%
\providecommand \@sanitize@url [0]{\catcode `\\12\catcode `\$12\catcode
  `\&12\catcode `\#12\catcode `\^12\catcode `\_12\catcode `\%12\relax}%
\providecommand \@@startlink[1]{}%
\providecommand \@@endlink[0]{}%
\providecommand \url  [0]{\begingroup\@sanitize@url \@url }%
\providecommand \@url [1]{\endgroup\@href {#1}{\urlprefix }}%
\providecommand \urlprefix  [0]{URL }%
\providecommand \Eprint [0]{\href }%
\providecommand \doibase [0]{http://dx.doi.org/}%
\providecommand \selectlanguage [0]{\@gobble}%
\providecommand \bibinfo  [0]{\@secondoftwo}%
\providecommand \bibfield  [0]{\@secondoftwo}%
\providecommand \translation [1]{[#1]}%
\providecommand \BibitemOpen [0]{}%
\providecommand \bibitemStop [0]{}%
\providecommand \bibitemNoStop [0]{.\EOS\space}%
\providecommand \EOS [0]{\spacefactor3000\relax}%
\providecommand \BibitemShut  [1]{\csname bibitem#1\endcsname}%
\let\auto@bib@innerbib\@empty
\bibitem [{\citenamefont {Madsen}\ \emph {et~al.}(2013)\citenamefont {Madsen},
  \citenamefont {Bergholtz},\ and\ \citenamefont {Brouwer}}]{Madsen}%
  \BibitemOpen
  \bibfield  {author} {\bibinfo {author} {\bibfnamefont {K.~A.}\ \bibnamefont
  {Madsen}}, \bibinfo {author} {\bibfnamefont {E.~J.}\ \bibnamefont
  {Bergholtz}}, \ and\ \bibinfo {author} {\bibfnamefont {P.~W.}\ \bibnamefont
  {Brouwer}},\ }\href@noop {} {\  (\bibinfo {year} {2013})},\ \Eprint
  {http://arxiv.org/abs/arXiv:1307.2577} {arXiv:1307.2577} \BibitemShut
  {NoStop}%
\bibitem [{\citenamefont {Kraus}\ \emph {et~al.}(2012)\citenamefont {Kraus},
  \citenamefont {Lahini}, \citenamefont {Ringel}, \citenamefont {Verbin},\ and\
  \citenamefont {Zilberberg}}]{us1}%
  \BibitemOpen
  \bibfield  {author} {\bibinfo {author} {\bibfnamefont {Y.~E.}\ \bibnamefont
  {Kraus}}, \bibinfo {author} {\bibfnamefont {Y.}~\bibnamefont {Lahini}},
  \bibinfo {author} {\bibfnamefont {Z.}~\bibnamefont {Ringel}}, \bibinfo
  {author} {\bibfnamefont {M.}~\bibnamefont {Verbin}}, \ and\ \bibinfo {author}
  {\bibfnamefont {O.}~\bibnamefont {Zilberberg}},\ }\href@noop {} {\bibfield
  {journal} {\bibinfo  {journal} {Phys. Rev. Lett.}\ }\textbf {\bibinfo
  {volume} {109}},\ \bibinfo {pages} {106402} (\bibinfo {year} {2012})},\
  \bibinfo {note} {and Supplemental Material}\BibitemShut {NoStop}%
\bibitem [{\citenamefont {Shinsei}\ \emph {et~al.}(2010)\citenamefont
  {Shinsei}, \citenamefont {Schnyder}, \citenamefont {Furusaki},\ and\
  \citenamefont {Ludwig}}]{LudwigDim}%
  \BibitemOpen
  \bibfield  {author} {\bibinfo {author} {\bibfnamefont {R.}~\bibnamefont
  {Shinsei}}, \bibinfo {author} {\bibfnamefont {A.~P.}\ \bibnamefont
  {Schnyder}}, \bibinfo {author} {\bibfnamefont {A.}~\bibnamefont {Furusaki}},
  \ and\ \bibinfo {author} {\bibfnamefont {A.~W.~W.}\ \bibnamefont {Ludwig}},\
  }\href@noop {} {\bibfield  {journal} {\bibinfo  {journal} {New J. Phys.}\
  }\textbf {\bibinfo {volume} {12}},\ \bibinfo {pages} {065010} (\bibinfo
  {year} {2010})}\BibitemShut {NoStop}%
\bibitem [{\citenamefont {Kraus}\ and\ \citenamefont {Zilberberg}(2012)}]{us2}%
  \BibitemOpen
  \bibfield  {author} {\bibinfo {author} {\bibfnamefont {Y.~E.}\ \bibnamefont
  {Kraus}}\ and\ \bibinfo {author} {\bibfnamefont {O.}~\bibnamefont
  {Zilberberg}},\ }\href@noop {} {\bibfield  {journal} {\bibinfo  {journal}
  {Phys. Rev. Lett.}\ }\textbf {\bibinfo {volume} {109}},\ \bibinfo {pages}
  {116404} (\bibinfo {year} {2012})}\BibitemShut {NoStop}%
\bibitem [{\citenamefont {Verbin}\ \emph {et~al.}(2013)\citenamefont {Verbin},
  \citenamefont {Zilberberg}, \citenamefont {Kraus}, \citenamefont {Lahini},\
  and\ \citenamefont {Silberberg}}]{us3}%
  \BibitemOpen
  \bibfield  {author} {\bibinfo {author} {\bibfnamefont {M.}~\bibnamefont
  {Verbin}}, \bibinfo {author} {\bibfnamefont {O.}~\bibnamefont {Zilberberg}},
  \bibinfo {author} {\bibfnamefont {Y.~E.}\ \bibnamefont {Kraus}}, \bibinfo
  {author} {\bibfnamefont {Y.}~\bibnamefont {Lahini}}, \ and\ \bibinfo {author}
  {\bibfnamefont {Y.}~\bibnamefont {Silberberg}},\ }\href@noop {} {\bibfield
  {journal} {\bibinfo  {journal} {Phys. Rev. Lett.}\ }\textbf {\bibinfo
  {volume} {110}},\ \bibinfo {pages} {076403} (\bibinfo {year}
  {2013})}\BibitemShut {NoStop}%
\bibitem [{\citenamefont {Moser}(1967)}]{Moser1967}%
  \BibitemOpen
  \bibfield  {author} {\bibinfo {author} {\bibfnamefont {J.}~\bibnamefont
  {Moser}},\ }\href@noop {} {\bibfield  {journal} {\bibinfo  {journal}
  {Mathematische Annalen}\ }\textbf {\bibinfo {volume} {169}},\ \bibinfo
  {pages} {136} (\bibinfo {year} {1967})}\BibitemShut {NoStop}%
\bibitem [{\citenamefont {Hale}(1969)}]{Hale1969}%
  \BibitemOpen
  \bibfield  {author} {\bibinfo {author} {\bibfnamefont {J.}~\bibnamefont
  {Hale}},\ }\href@noop {} {\emph {\bibinfo {title} {Ordinary differential
  equations}}},\ Pure and applied mathematics\ (\bibinfo  {publisher}
  {Wiley-Interscience},\ \bibinfo {year} {1969})\ p.\ \bibinfo {pages}
  {355}\BibitemShut {NoStop}%
\bibitem [{\citenamefont {Thouless}(1983)}]{Thouless_pump}%
  \BibitemOpen
  \bibfield  {author} {\bibinfo {author} {\bibfnamefont {D.~J.}\ \bibnamefont
  {Thouless}},\ }\href@noop {} {\bibfield  {journal} {\bibinfo  {journal}
  {Phys. Rev. B}\ }\textbf {\bibinfo {volume} {27}},\ \bibinfo {pages} {6083}
  (\bibinfo {year} {1983})}\BibitemShut {NoStop}%
\bibitem [{\citenamefont {Ringel}\ \emph {et~al.}(2012)\citenamefont {Ringel},
  \citenamefont {Kraus},\ and\ \citenamefont {Stern}}]{us_WTI}%
  \BibitemOpen
  \bibfield  {author} {\bibinfo {author} {\bibfnamefont {Z.}~\bibnamefont
  {Ringel}}, \bibinfo {author} {\bibfnamefont {Y.~E.}\ \bibnamefont {Kraus}}, \
  and\ \bibinfo {author} {\bibfnamefont {A.}~\bibnamefont {Stern}},\
  }\href@noop {} {\bibfield  {journal} {\bibinfo  {journal} {Phys. Rev. B}\
  }\textbf {\bibinfo {volume} {86}},\ \bibinfo {pages} {045102} (\bibinfo
  {year} {2012})}\BibitemShut {NoStop}%
\bibitem [{\citenamefont {Fu}(2011)}]{Fu_CTI}%
  \BibitemOpen
  \bibfield  {author} {\bibinfo {author} {\bibfnamefont {L.}~\bibnamefont
  {Fu}},\ }\href@noop {} {\bibfield  {journal} {\bibinfo  {journal} {Phys. Rev.
  Lett.}\ }\textbf {\bibinfo {volume} {106}},\ \bibinfo {pages} {106802}
  (\bibinfo {year} {2011})}\BibitemShut {NoStop}%
\bibitem [{\citenamefont {Janot}(1994)}]{QC_Janot}%
  \BibitemOpen
  \bibfield  {author} {\bibinfo {author} {\bibfnamefont {C.}~\bibnamefont
  {Janot}},\ }\href@noop {} {\emph {\bibinfo {title} {Quasicrystals}}},\
  \bibinfo {edition} {2nd}\ ed.\ (\bibinfo  {publisher} {Clarendon},\ \bibinfo
  {address} {Oxford},\ \bibinfo {year} {1994})\BibitemShut {NoStop}%
\bibitem [{\citenamefont {Aubry}\ and\ \citenamefont {Andr\'{e}}(1980)}]{AA}%
  \BibitemOpen
  \bibfield  {author} {\bibinfo {author} {\bibfnamefont {S.}~\bibnamefont
  {Aubry}}\ and\ \bibinfo {author} {\bibfnamefont {G.}~\bibnamefont
  {Andr\'{e}}},\ }\href@noop {} {\bibfield  {journal} {\bibinfo  {journal}
  {Ann. Isr. Phys. Soc.}\ }\textbf {\bibinfo {volume} {3}},\ \bibinfo {pages}
  {133} (\bibinfo {year} {1980})}\BibitemShut {NoStop}%
\bibitem [{\citenamefont {Roati}\ \emph {et~al.}(2008)\citenamefont {Roati}
  \emph {et~al.}}]{Roati08}%
  \BibitemOpen
  \bibfield  {author} {\bibinfo {author} {\bibfnamefont {G.}~\bibnamefont
  {Roati}} \emph {et~al.},\ }\href@noop {} {\bibfield  {journal} {\bibinfo
  {journal} {Nature}\ }\textbf {\bibinfo {volume} {453}},\ \bibinfo {pages}
  {895} (\bibinfo {year} {2008})}\BibitemShut {NoStop}%
\bibitem [{\citenamefont {Lahini}\ \emph {et~al.}(2009)\citenamefont {Lahini},
  \citenamefont {Pugatch}, \citenamefont {Pozzi}, \citenamefont {Sorel},
  \citenamefont {Morandotti}, \citenamefont {Davidson},\ and\ \citenamefont
  {Silberberg}}]{Lahini}%
  \BibitemOpen
  \bibfield  {author} {\bibinfo {author} {\bibfnamefont {Y.}~\bibnamefont
  {Lahini}}, \bibinfo {author} {\bibfnamefont {R.}~\bibnamefont {Pugatch}},
  \bibinfo {author} {\bibfnamefont {F.}~\bibnamefont {Pozzi}}, \bibinfo
  {author} {\bibfnamefont {M.}~\bibnamefont {Sorel}}, \bibinfo {author}
  {\bibfnamefont {R.}~\bibnamefont {Morandotti}}, \bibinfo {author}
  {\bibfnamefont {N.}~\bibnamefont {Davidson}}, \ and\ \bibinfo {author}
  {\bibfnamefont {Y.}~\bibnamefont {Silberberg}},\ }\href@noop {} {\bibfield
  {journal} {\bibinfo  {journal} {Phys. Rev. Lett.}\ }\textbf {\bibinfo
  {volume} {103}},\ \bibinfo {pages} {013901} (\bibinfo {year}
  {2009})}\BibitemShut {NoStop}%
\bibitem [{\citenamefont {Ezawa}\ \emph {et~al.}(2013)\citenamefont {Ezawa},
  \citenamefont {Tanaka},\ and\ \citenamefont {Nagaosa}}]{Nagaosa:2013}%
  \BibitemOpen
  \bibfield  {author} {\bibinfo {author} {\bibfnamefont {M.}~\bibnamefont
  {Ezawa}}, \bibinfo {author} {\bibfnamefont {Y.}~\bibnamefont {Tanaka}}, \
  and\ \bibinfo {author} {\bibfnamefont {N.}~\bibnamefont {Nagaosa}},\
  }\href@noop {} {\  (\bibinfo {year} {2013})},\ \bibinfo {note} {has a similar
  discussion},\ \Eprint {http://arxiv.org/abs/arXiv:1307.7347}
  {arXiv:1307.7347} \BibitemShut {NoStop}%
\bibitem [{\citenamefont {Kraus}\ \emph {et~al.}(2013)\citenamefont {Kraus},
  \citenamefont {Ringel},\ and\ \citenamefont {Zilberberg}}]{us4}%
  \BibitemOpen
  \bibfield  {author} {\bibinfo {author} {\bibfnamefont {Y.~E.}\ \bibnamefont
  {Kraus}}, \bibinfo {author} {\bibfnamefont {Z.}~\bibnamefont {Ringel}}, \
  and\ \bibinfo {author} {\bibfnamefont {O.}~\bibnamefont {Zilberberg}},\
  }\href@noop {} {\  (\bibinfo {year} {2013})},\ \Eprint
  {http://arxiv.org/abs/arXiv:1302.2647} {arXiv:1302.2647} \BibitemShut
  {NoStop}%
\end{thebibliography}

%

\end{document}